\documentclass[aps,12pt,showpacs,preprint,groupedaddress,floatfix]{revtex4}
\usepackage{graphicx}
\usepackage{dcolumn}
\usepackage{bm}
\usepackage{amssymb}

\begin{document}

\title{Microscopic description of nuclear quantum phase transitions}
\author{T. Nik\v si\' c}
\author{D. Vretenar}
\affiliation{Physics Department, University of Zagreb, 
Bijeni\v cka 32, Zagreb, Croatia}
\author{G. A. Lalazissis}
\affiliation{Department of Theoretical Physics, Aristotle University of
Thessaloniki, GR-54124, Greece}
\author{P. Ring}
\affiliation{Physik-Department der Technischen Universit\"at M\"unchen,
D-85748 Garching,
Germany}
\date{\today}
\begin{abstract}
The relativistic mean-field framework,
extended to include correlations related to 
restoration of broken symmetries and to fluctuations of
the quadrupole deformation, is applied to a study of 
shape transitions in Nd isotopes. It is demonstrated that the 
microscopic self-consistent approach, based on global 
effective interactions, can describe not only general features 
of transitions between spherical and deformed nuclei, but 
also the singular properties of excitation spectra and 
transition rates at the critical point of quantum shape 
phase transition.
\end{abstract}

\maketitle

Quantum phase transitions (QPT) between competing ground-state phases can 
be induced by variation of a non-thermal control parameter at zero temperature.
In the case of atomic nuclei first- and second-order QPT can occur 
between systems characterized by different ground-state shapes. Nuclear shape 
phase transitions have been the subject of a number of recent theoretical and 
experimental studies, pioneered by Iachello \cite{FI.00}, Casten and 
Zamfir \cite{CZ.00}. These transitions can be 
described in the geometric framework in terms of a Bohr Hamiltonian for 
shape variables, and related to the concept of critical symmetries which 
provide parameter independent 
predictions for excitation spectra and electric quadrupole (E2)
 transition rates for nuclei at the phase transition 
point \cite{FI.00}. Alternatively, in the algebraic approach 
different shapes coincide with particular dynamic symmetries of some algebraic 
structure, and the QPT occurs when these symmetries are broken in 
a specific way \cite{FI.00,CJ.00}. In both approaches, geometric and algebraic, 
the description of QPT is based on model specific Hamiltonians which by 
construction describe shape changes. A shape phase transition is then 
accessed by variation of a control parameter. Analytic solutions of the 
eigenvalue problem at the critical point are associated with zeros of 
special functions, corresponding to a particular (critical) symmetry of the 
Hamiltonian.

Initially two critical point symmetries were introduced: E(5) 
and X(5) \cite{FI.00},  which correspond to a second-order 
QPT between spherical and 
$\gamma$-soft shapes, and 
a first-order QPT between spherical and 
axially deformed shapes, respectively. The O(6) 
symmetry limit of the interacting boson model 
has also been identified as another critical point symmetry, 
for the transition between prolate and oblate 
SU(3) shapes \cite{Jan.01}. The study of Ref.~\cite{Jan.02} 
has shown that the isolated point of second-order phase 
transition, labeled E(5), represents a triple point where lines 
of first-order phase transition meet. QPT as a function of the 
nucleon number has become an important current topic 
in nuclear structure, and encompasses a wide range of 
subjects, from dynamical symmetries and Landau theory, 
to the concept of order and chaos in nuclear spectra. For 
recent reviews we refer 
the reader to Refs.~\cite{JC.05,Rick.06}.

In this work we address the important question whether nuclear QPT can 
also be described in a general microscopic framework, based on effective 
nucleon-nucleon interactions or universal energy density functionals, 
that provides a unified description of bulk properties (masses, density 
distributions, radii) and excitation spectra. A variety of structure phenomena 
have been described with self-consistent mean-field (SCMF)
models based on the Gogny interaction, the Skyrme energy 
functional, and the relativistic meson-exchange effective Lagrangian 
\cite{VALR.05}. Important advantages of the mean-field approach include 
the use of global effective nuclear interactions, the treatment of arbitrarily 
heavy systems, and the intuitive picture of intrinsic shapes. 

In several recent studies \cite{Meng.05,FBL.06}, the SCMF
framework has been employed in calculations of potential 
energy curves (PECs) as functions of the quadrupole deformation, for  
isotopic chains in which the occurrence of critical point symmetries 
had been predicted. The resulting PECs display shape transitions from 
spherical to deformed configurations. It was shown that 
particular isotopes exhibit relatively flat PECs over an extended 
range of the deformation parameter, and this has been interpreted 
as signature of certain types of critical point symmetries. A simple 
mean-field approach, however, cannot be used for a quantitative 
analysis of critical point nuclei. Flat PECs are one of 
the characteristics of critical point symmetries, but PECs alone do 
not single out a specific isotope as being the critical one. The concept 
of shape phase transition and related critical point symmetry includes 
analytic expressions for observables: 
excitation energies and transition rates. Thus in order to attribute 
a critical point symmetry to a specific nucleus, one must be able to calculate the 
ratios of excitation energies and B(E2) rates, and this is not possible on the 
mean-field level.

For a quantitative description of nuclei with soft PECs
it is necessary to explicitly consider correlation effects beyond the mean-field 
level: the restoration of rotational symmetry and fluctuations of the quadrupole
deformation. This allows the calculation of spectra and transition rates, and 
therefore provides the basis for a quantitative prediction of QPT.
In addition, in open-shell nuclei projection on particle number is crucial whenever 
the number of correlated pairs becomes small and the density of levels is low, a 
situation typical for the description of phenomena related to the evolution of shell 
structure. In order to single out the critical point isotope, it is thus necessary 
to perform the projection on eigenstates of the particle number operators.
 
In two recent articles \cite{NVR.06} we have developed a 
model which extends the self-consistent relativistic mean-field
(RMF) approach
to include correlations related to restoration of broken symmetries 
and to fluctuations of collective variables. The model uses the 
generator coordinate method (GCM) to perform configuration 
mixing of angular-momentum and
particle-number projected many-body wave functions,
generated from the solutions of
the RMF + Lipkin-Nogami
BCS equations, with a constraint on the mass quadrupole moment.
A point-coupling nucleon-nucleon effective interaction 
is used in the particle-hole
channel, and a density-independent $\delta$-interaction in the
pairing channel.

In the present study we apply this model to the 
description of shape transitions and address the following question:
can a universal density functional, with parameters adjusted to 
global ground-state properties (masses, radii), at the same time reproduce 
the singular behavior of excitation spectra at the critical point of shape 
phase transition?

In the current version the model is restricted to  
axially symmetric shapes. We will thus consider transitions between 
spherical and axially deformed shapes in the chain of Nd isotopes. 
In the language of the interacting boson model this is a 
transition between the U(5) and SU(3) dynamical symmetries, and in 
Ref.~\cite{FI.00} it has been shown that along this path 
a first-order shape phase transition occurs, 
associated with the X(5) critical symmetry. An approximate 
solution in terms of zeros of Bessel functions of irrational order 
was presented for the particular case in which the $\beta$ and 
$\gamma$ degrees of freedom are decoupled and only the 
$\beta$-term is retained in the transition operator. Evidence for 
the empirical realization of X(5) critical symmetry was first
reported for $^{152}$Sm and other $N=90$ isotones in 
Ref.~\cite{CZ.00}. By comparing the experimental low-spin level 
scheme and reduced transition probabilities with parameter-free 
X(5) predictions, in Ref.~\cite{Rei.02} it has been shown that 
$^{150}$Nd presents a very good case of X(5) critical symmetry. 
This symmetry has been studied in a number of recent papers, both in 
the geometric and algebraic approach and, in particular, the 
dynamics at the critical point of a general first-order quantum phase 
transition has been analyzed from an algebraic 
perspective in Ref.~\cite{Ami.06}.  An exactly separable $\gamma$-rigid 
version (with $\gamma = 0$) of the X(5) symmetry has been 
constructed  recently\cite{Bon.06}. 

In the following we report self-consistent GCM calculations 
based on the relativistic point-coupling effective interaction 
PC-F1 \cite{BMM.02}, which has been adjusted to ground-state 
observables of spherical nuclei, and tested
in the analysis of the equation of state of symmetric nuclear matter and 
neutron matter, binding energies and form-factors, and 
ground-state properties of several isotopic and isotonic chains. 
In addition to the mean-field 
potential, pairing correlations are described 
in the Lipkin-Nogami (LN) BCS approximation. A density-independent 
$\delta$-force is used as the effective interaction in the particle-particle 
channel. As explained in Ref.~\cite{NVR.06}, 
the proton and neutron pairing strengths that were 
determined simultaneously with the PC-F1 parameters, have to be 
reduced by $\approx$ 10\% so that the projected average pairing 
gaps reproduce the BCS pairing gaps: 
$V_p = -260$ MeV and $V_n = -285$ MeV. Ref.~\cite{NVR.06} includes 
a detailed description of the relativistic GCM model 
with angular-momentum and particle-number projections.
\begin{figure}
\includegraphics[scale=0.55]{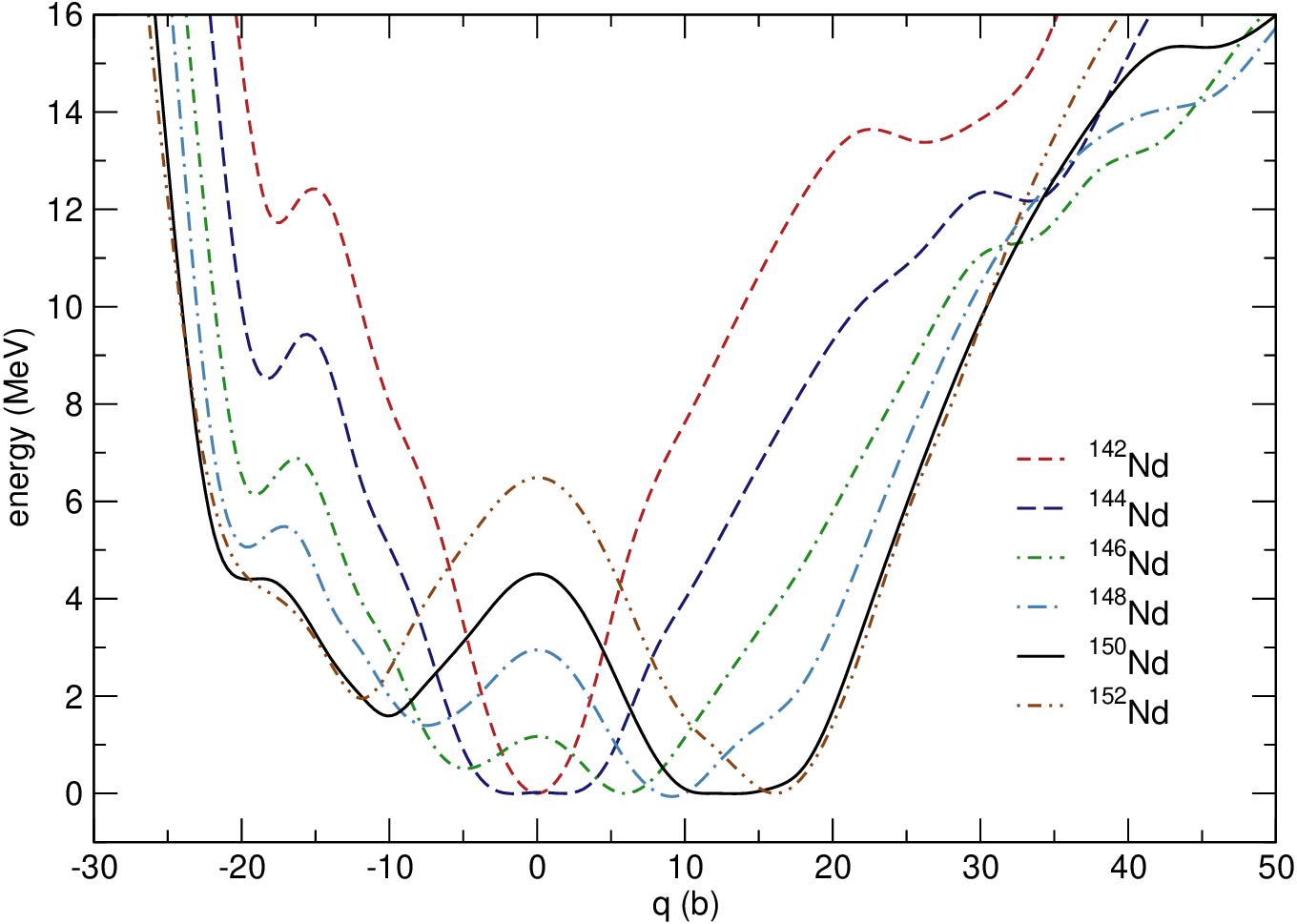}
\caption{(Color online) Self-consistent RMF  
binding energy curves of $^{142-152}$Nd, as 
functions of the mass quadrupole moment.}
\label{figA}
\end{figure}

For an axially deformed nucleus the energy surface 
as function of deformation is obtained by imposing a
constraint on the mass quadrupole moment.
The mean-field breaks rotational symmetry, and the particle
number is only approximately restored by the Lipkin-Nogami procedure. 
In order to be able to compare model predictions with data, states 
with good angular momentum and particle number are projected 
from the mean-field + LN BCS solutions. For each value of the angular 
momentum the solution of the Hill-Wheller eigenvalue equations determines
the excitation spectrum.  In Fig.~\ref{figA} we plot the
self-consistent RMF + LN 
BCS binding energy curves of $^{142-152}$Nd, as functions of the 
mass quadrupole moment. The PECs display a gradual transition between 
the spherical $^{142}$Nd and strongly prolate deformed $^{152}$Nd. Of 
particular interest here is the PEC of $^{150}$Nd which exhibits a wide 
flat minimum on the prolate side ($\gamma = 0^{\circ}$), with an 
additional miminum at $\approx$ 1.8 MeV excitation energy and 
oblate ($\gamma = 60^{\circ}$) deformation. 
The two minima are separated by a potential barrier of $\approx 4.5$ MeV. 
One notes the similarity between the PEC of $^{150}$Nd shown 
in Fig.~\ref{figA}, and the projections on the $\gamma = 0^{\circ}$ (prolate) 
and $\gamma = 60^{\circ}$ (oblate) axes of the original X(5) potential 
considered by Iachello in Ref.~\cite{FI.00} (square well in 
the variable $\beta$, and harmonic oscillator potential in $\gamma$).
In Ref.~\cite{PG.04} the X(5) solution has been generalized to the 
transition between X(5) and the rigid-rotor limit by 
considering an infinite square well over a confined range 
of values $\beta_M > \beta_m \geq 0$.
\begin{figure}
\includegraphics[scale=0.65]{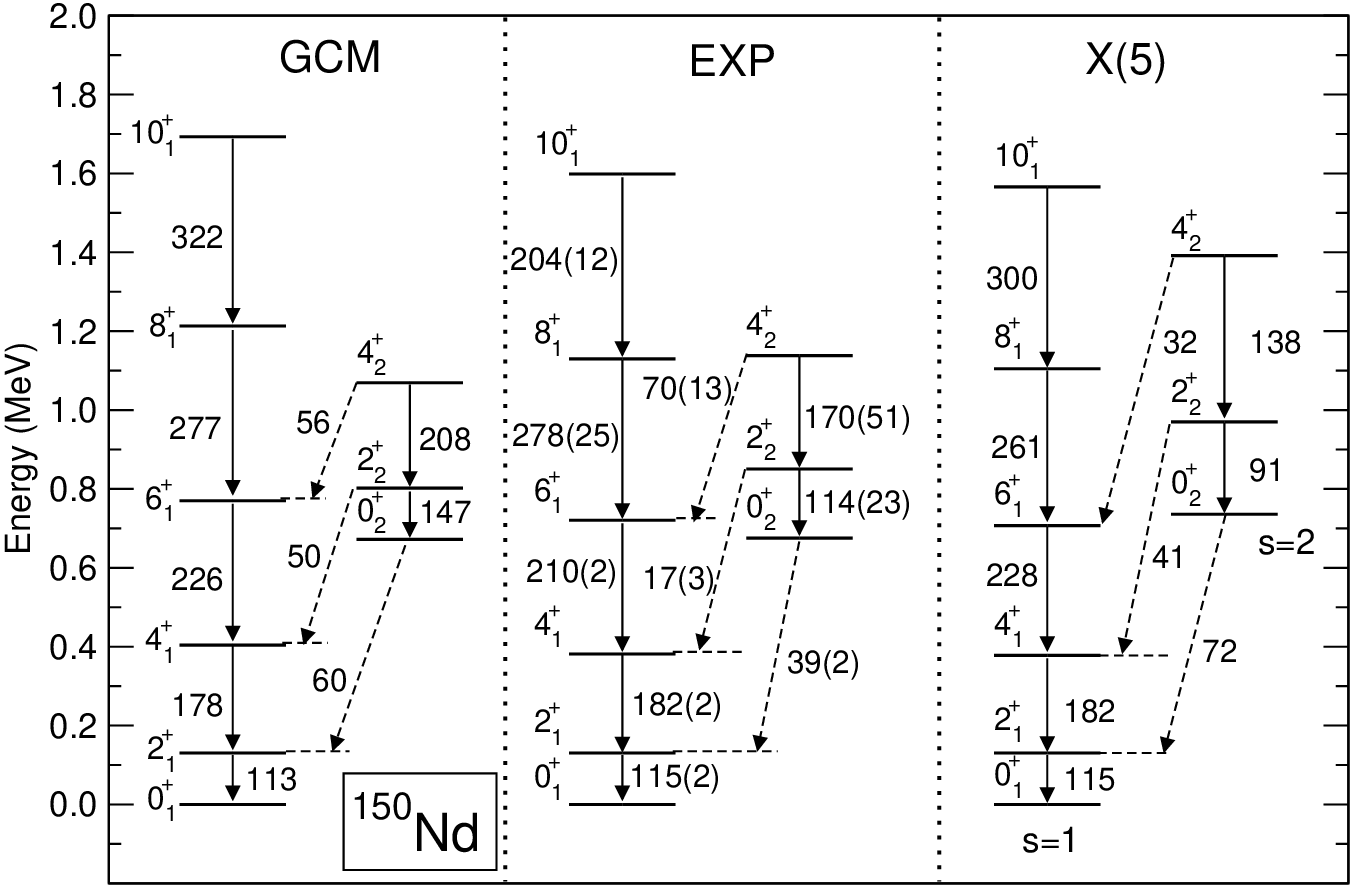}
\caption{The particle-number projected GCM spectrum of  $^{150}$Nd 
(left), compared with the data \cite{Rei.02} (middle), and the 
X(5)-symmetry predictions (right) for the excitation energies, intraband 
and interband B(E2) values (in Weisskopf units) of the ground-state 
($s=1$) and $\beta_1$ ($s=2$) bands. 
The theoretical spectra are normalized to the experimental energy of the
state $2^+_1$, and the X(5) transition strengths are normalized
to the experimental B(E2; $2^+_1 \to 0^+_1$).}
\label{figB}
\end{figure}
\begin{figure}
\includegraphics[scale=0.65]{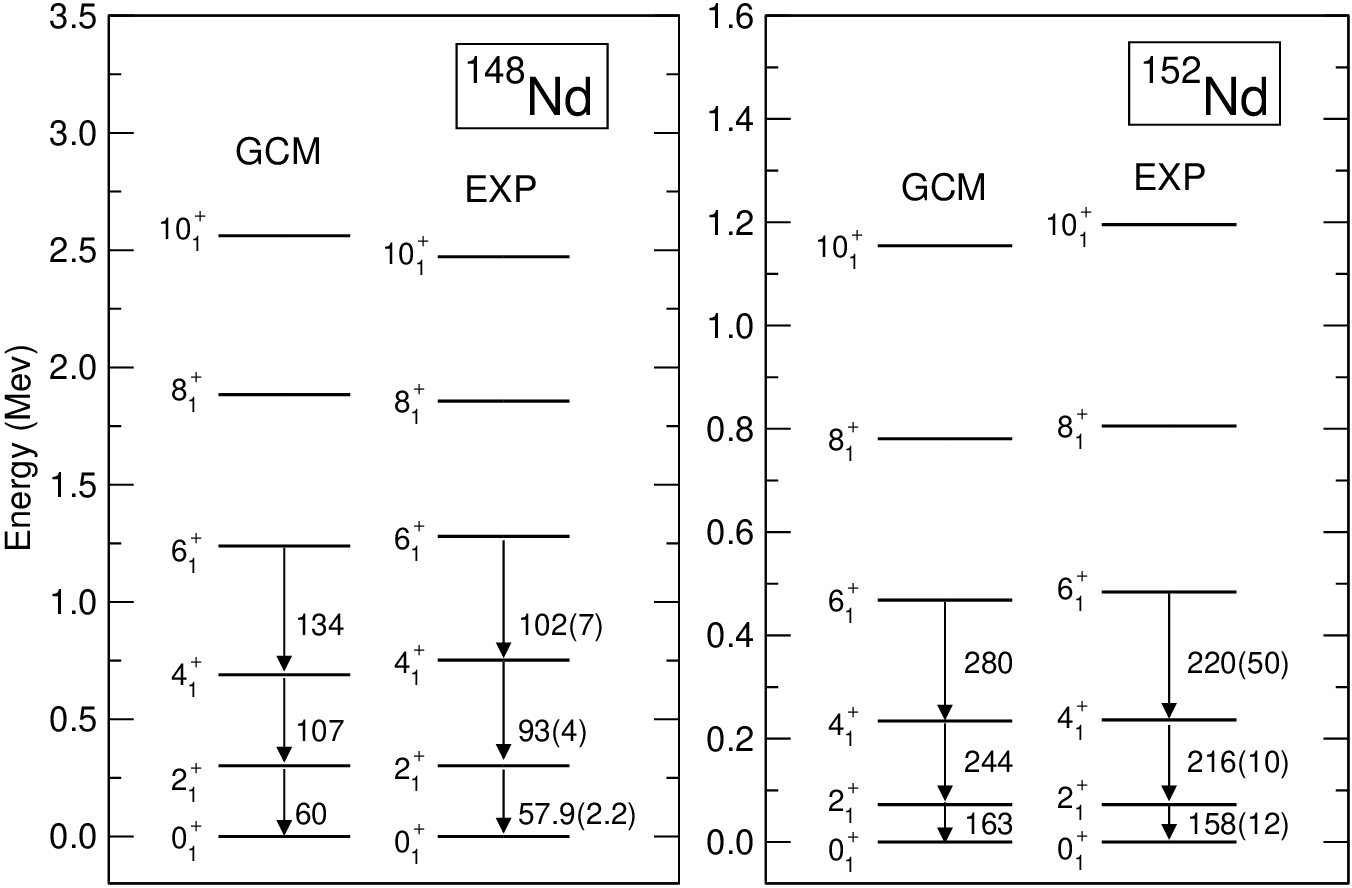}
\caption{The particle-number projected GCM results for the 
excitation energies and intraband B(E2) 
values (in Weisskopf units) of the ground-state bands of 
$^{148}$Nd (left) and $^{152}$Nd 
(right),  are shown in comparison with the data \cite{148.dat,152.dat}.}
\label{figC}
\end{figure}

Correlation effects related to the restoration of broken symmetries and
to fluctuations of collective coordinates are taken into account by
performing configuration mixing calculations of projected states. 
For $^{150}$Nd we have thus solved the GCM equations in 
the basis of constrained mean-field + LN BCS Slater determinants, 
projected on angular momentum and particle number. The GCM results 
for the two lowest bands are compared in Fig.~\ref{figB} with the available 
data \cite{Rei.02}, and with the X(5)-symmetry predictions for the excitation 
energies, intraband B(E2) values (in Weisskopf units) of the 
ground-state ($s=1$) and $\beta_1$ ($s=2$) bands, 
and interband transitions between the ($s=2$) and ($s=1$) bands.
To facilitate the comparison with 
the X(5) spectrum, which corresponds to the solution around  
$\gamma = 0^{\circ}$ \cite{FI.00}, the GCM results in Fig.~\ref{figB} 
have been obtained by performing configuration 
mixing calculations only on the prolate side (positive values 
of the quadrupole moment in Fig.~\ref{figA}). 
The theoretical spectra are normalized to the experimental energy of the
state $2^+_1$ and, in addition, the X(5) transition strengths are normalized
to the experimental B(E2; $2^+_1 \to 0^+_1$). In the mean-field plus 
GCM approach the transition rates are calculated in the full configuration space 
using bare charges, and the BE(2) values can be directly compared with data. 
We note the excellent agreement of the GCM spectrum both with the data and 
with the X(5)-symmetry predictions. 

When the additional mixing with oblate ($\gamma = 60^{\circ}$) 
configurations is allowed in the GCM spectrum 
of  $^{150}$Nd, a visible effect 
is obtained only for the ground state $0^+_1$, which gets 
lowered in energy by $\approx 100$ keV. The other states in the lowest two 
bands, and the corresponding E2 transition rates, are not affected by the mixing 
with $\gamma = 60^{\circ}$ configurations.
In fact, since the basis of deformed states does not include the full 
range of $\gamma$-values ($0^{\circ} \leq \gamma \leq 60^{\circ}$), 
configuration mixing must be performed only on the prolate side 
in order to remain close to the phase-transitional region in which 
X(5) occurs. The spectrum resulting from a 
mixing of pure prolate ($\gamma = 0^{\circ}$) 
and oblate ($\gamma = 60^{\circ}$) configurations corresponds 
to a location on the hypotenuse of the right-angled extended Casten 
triangle \cite{Jan.01,Jan.02,Rick.06}, opposite to the side along 
which the X(5) transition takes place.
\begin{figure}
\includegraphics[scale=0.55]{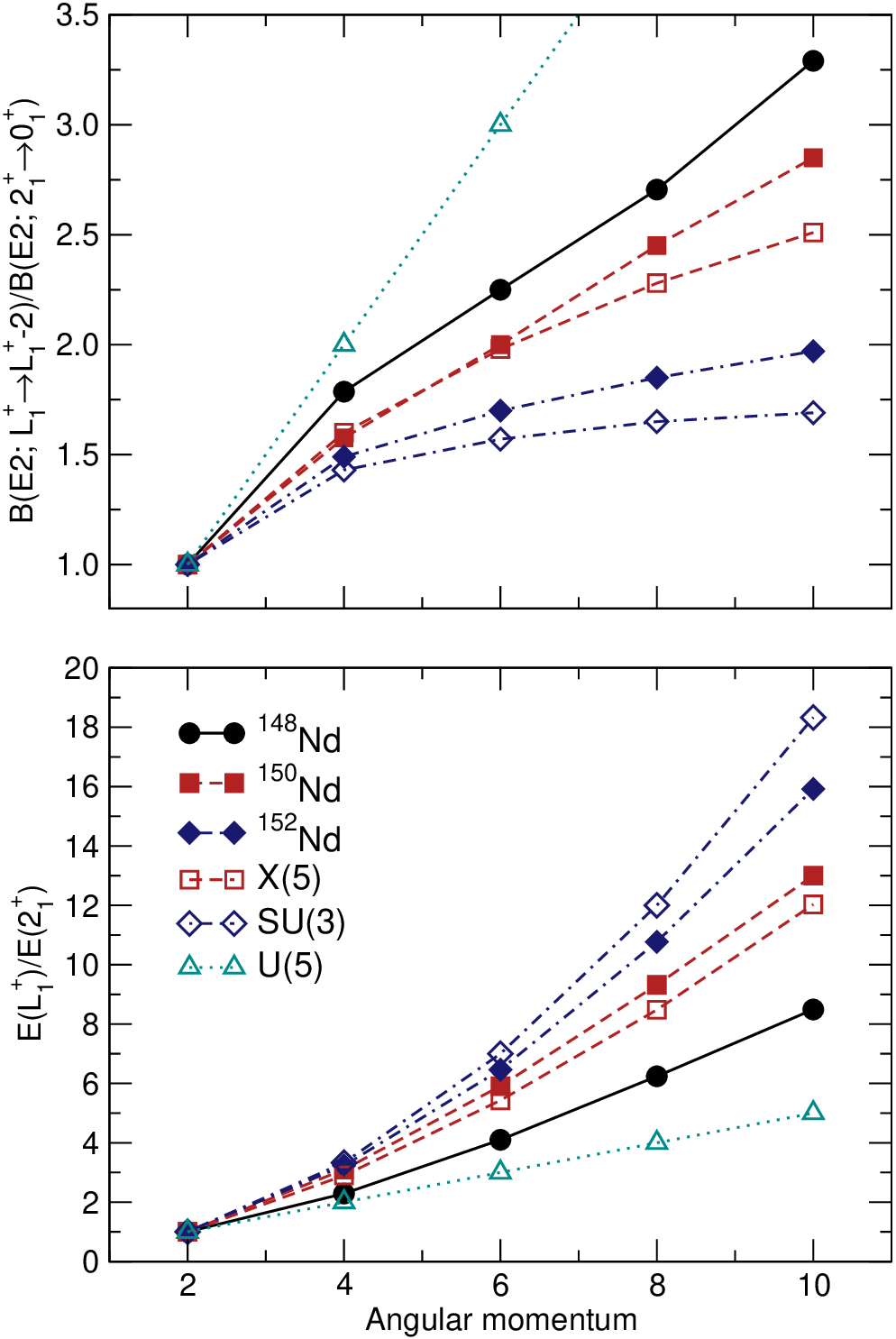}
\caption{(Color online) B(E2; $L \to L-2$) values (upper panel) and 
excitation energies (lower panel) for the yrast states in $^{148}$Nd, 
$^{150}$Nd, and $^{152}$Nd, calculated with the GCM and compared 
with those predicted by the X(5), SU(3) and U(5) symmetries (open symbols).}
\label{figD}
\end{figure}

In order to demonstrate that the self-consistent GCM calculation 
based on the effective interaction 
PC-F1 predicts the shape phase transition precisely at the 
isotope $^{150}$Nd, we also need to consider the 
neighboring nuclei $^{148}$Nd and $^{152}$Nd. In Fig.~\ref{figC} 
the GCM results for the 
excitation energies and B(E2) 
values (in Weisskopf units) of the ground-state bands of 
$^{148}$Nd and $^{152}$Nd 
are shown in comparison with available data \cite{148.dat,152.dat}.
The agreement with experiment is very good and of course we notice 
that the calculated bands and transition rates are nothing
like the X(5) spectrum of $^{150}$Nd. The phase transition
is further illustrated in Fig.~\ref{figD} where, 
for the yrast states of $^{148}$Nd, $^{150}$Nd and $^{152}$Nd, 
we compare the B(E2; $L \to L-2$) values 
and excitation energies calculated in the relativistic GCM 
model, with the corresponding values predicted 
by the U(5),  X(5), and SU(3) 
symmetries.  The spectrum of $^{148}$Nd differs from
these symmetry limits, as can already be inferred
from the calculated PES in Fig.~\ref{figA}. 
The GCM E2 rates and excitation energies for $^{150}$Nd closely follow   
those calculated from analytic expressions corresponding to the X(5)
critical symmetry, whereas the yrast states of  $^{152}$Nd 
agree with the prediction of the SU(3) symmetry limit. 
The important result of the present analysis is that the X(5) critical 
symmetry, and therefore the shape phase transition, arises in the 
calculated spectrum of $^{150}$Nd as a result of quadrupole shape 
fluctuations. Namely, at each point in the flat prolate minimum shown 
in Fig.~\ref{figA}, the yrast spectrum obtained by angular momentum 
projection from the mean-field PEC corresponds to a rotor, 
i.e. to the SU(3) symmetry limit. Only when quadrupole fluctuations 
are taken into account by GCM configuration mixing 
calculations, the resulting spectrum approaches the X(5) critical symmetry 
(Figs.~\ref{figB} and \ref{figD}).

In conclusion, we have demonstrated that the microscopic self-consistent 
mean-field framework, based on a universal energy density functional 
adjusted to nuclear ground-state properties, and extended to take into 
account correlations related to symmetry restoration and fluctuations 
of collective variables, describes not only general features of shape 
transitions, but also the singular behavior of the excitation spectra and 
transition rates at the critical point of a quantum shape phase transition.

\bigskip
\leftline{\bf ACKNOWLEDGMENTS} 
We thank F. Iachello and D. Bonatsos for useful discussions. 
This work has been supported in part
by MZOS - 
project 1191005-1010, by Pythagoras II - EPEAK II and EU project 80861,
and by BMBF - project 06 MT 246. 
\bigskip

\end{document}